\documentclass[fleqn,usenatbib]{mnras}

\usepackage[T1]{fontenc}

\DeclareRobustCommand{\VAN}[3]{#2}
\let\VANthebibliography\thebibliography
\def\thebibliography{\DeclareRobustCommand{\VAN}[3]{##3}\VANthebibliography}

\usepackage{graphicx}	
\usepackage{amsmath}	

\usepackage{newtxtext,newtxmath}

\title[Solenoidal turbulence in the Brick]{Kinematics of Galactic Centre clouds shaped by shear-seeded solenoidal turbulence}

\author[M. A. Petkova et al.]{Maya~A.~Petkova,$^{1,2}$\thanks{E-mail: maya.petkova@chalmers.se (MAP)}
J.~M.~Diederik~Kruijssen,$^{3,4}$
Jonathan~D.~Henshaw,$^{5,6}$
Steven~N.~Longmore,$^{5,4}$
\newauthor
Simon~C.~O.~Glover,$^{7}$
Mattia~C.~Sormani,$^{7}$
Lucia~Armillotta,$^{8}$
Ashley~T.~Barnes,$^{9}$
Ralf~S.~Klessen,$^{7,10}$
\newauthor
Francisco~Nogueras-Lara,$^{11}$
Robin~G.~Tress,$^{12}$
Jairo Armijos-Abenda{\~n}o,$^{13,14}$
Laura Colzi,$^{15}$
\newauthor
Christoph Federrath,$^{16,17}$
Pablo Garc{\'i}a,$^{18,19}$
Adam Ginsburg,$^{20}$
Christian Henkel,$^{21,22}$
Sergio Mart{\'i}n,$^{23,24}$
\newauthor
Denise Riquelme$^{21,25,26}$
and V{\'i}ctor M. Rivilla$^{15}$
\\
$^{1}$Astronomisches Rechen-Institut, Zentrum f{\"u}r Astronomie der Universit{\"a}t Heidelberg, M{\"o}nchhofstr 12-14, D-69120 Heidelberg, Germany\\
$^{2}$Space, Earth and Environment Department, Chalmers University of Technology, SE-412 96 Gothenburg, Sweden\\
$^{3}$Technical University of Munich, School of Engineering and Design, Department of Aerospace and Geodesy, Chair of Remote Sensing Technology, Arcisstr. 21, 80333 Munich, Germany\\
$^{4}$Cosmic Origins Of Life (COOL) Research DAO, coolresearch.io\\
$^{5}$Astrophysics Research Institute, Liverpool John Moores University, IC2, Liverpool Science Park, 146 Brownlow Hill, Liverpool L3 5RF, UK\\
$^{6}$Max-Planck-Institut f{\"u}r Astronomie, K{\"o}nigstuhl 17, D-69117, Heidelberg, Germany\\
$^{7}$Universit{\"a}t Heidelberg, Zentrum f{\"u}r Astronomie, Institut f{\"u}r Theoretische Astrophysik, Albert-Ueberle-Str 2, D-69120 Heidelberg, Germany\\
$^{8}$Department of Astrophysical Sciences, Princeton University, Princeton, NJ 08544, USA\\
$^{9}$Institut f{\"u}r Astronomie, Universit{\"a}t Bonn, Auf dem H{\"u}gel 71, 53121, Bonn, DE\\
$^{10}$ Universit\"{a}t Heidelberg, Interdisziplin\"{a}res Zentrum f\"{u}r Wissenschaftliches Rechnen, Im Neuenheimer Feld 205, D-69120 Heidelberg, Germany\\
$^{11}$ European Southern Observatory, Karl-Schwarzschild-Strasse 2, 85748 Garching bei M{\"u}nchen, Germany\\
$^{12}$ Institute of Physics, Laboratory for galaxy evolution and spectral modelling, EPFL, Observatoire de Sauverny, Chemin Pegais 51, 1290 Versoix, Switzerland\\
$^{13}$ School of Physics and Astronomy, Cardiff University, The Parade, Cardiff CF24 3AA, UK\\
$^{14}$ Observatorio Astron{\'o}mico de Quito, Escuela Polit{\'e}cnica Nacional, 170136, Quito, Ecuador\\
$^{15}$ Centro de Astrobiolog{\'i}a (CAB), CSIC-INTA, Ctra. de Ajalvir Km. 4, 28850, Torrej{\'o}n de Ardoz, Madrid, Spain\\
$^{16}$ Research School of Astronomy and Astrophysics, Australian National University, Canberra, ACT~2611, Australia\\
$^{17}$ Australian Research Council Centre of Excellence in All Sky Astrophysics (ASTRO3D), Canberra, ACT~2611, Australia\\
$^{18}$ Instituto de Astronom\'{i}a, Universidad Cat\'{o}lica del Norte, Av. Angamos 0610, Antofagasta 1270709, Chile\\
$^{19}$ Chinese Academy of Sciences South America Center for Astronomy, National Astronomical Observatories, CAS, Beijing 100101, China\\
$^{20}$ Department of Astronomy, University of Florida 211 Bryant Space Science Center P.O Box 112055, Gainesville, FL 32611-2055 USA\\
$^{21}$ Max-Planck-Institut f{\"u}r Radioastronomie, Auf dem H{\"u}gel 69, 53121 Bonn, Germany\\
$^{22}$ Astronomy Department, Faculty of Science, King Abdulaziz University, P. O. Box 80203, Jeddah 21589, Saudi Arabia\\
$^{23}$ European Southern Observatory, Alonso de C{\'o}rdova, 3107, Vitacura, Santiago, 763-0355, Chile\\
$^{24}$ Joint ALMA Observatory, Alonso de C{\'o}rdova, 3107, Vitacura, Santiago, 763-0355, Chile\\
$^{25}$ Departamento de Astronom\'ia, Universidad de La Serena, Av. Cisternas 1200, La Serena, Chile\\
$^{26}$ Instituto Multidisciplinario de Investigaci\'on y Postgrado, Universidad de La Serena, Ra\'ul Bitr\'an 1305, La Serena, Chile
}

\date{Accepted XXX. Received YYY; in original form ZZZ}

\pubyear{2023}

\begin{document}
\label{firstpage}
\pagerange{\pageref{firstpage}--\pageref{lastpage}}
\maketitle

\begin{abstract}
The Central Molecular Zone (CMZ; the central $\sim 500$~pc of the Galaxy) is a kinematically unusual environment relative to the Galactic disc, with high velocity dispersions and a steep size-linewidth relation of the molecular clouds. In addition, the CMZ region has a significantly lower star formation rate (SFR) than expected by its large amount of dense gas. An important factor in explaining the low SFR is the turbulent state of the star-forming gas, which seems to be dominated by rotational modes. However, the turbulence driving mechanism remains unclear. 
In this work, we investigate how the Galactic gravitational potential affects the turbulence in CMZ clouds. 
We focus on the CMZ cloud G0.253+0.016 (`the Brick’), which is very quiescent and unlikely to be kinematically dominated by stellar feedback. We demonstrate that several kinematic properties of the Brick arise naturally in a cloud-scale hydrodynamics simulation that takes into account the Galactic gravitational potential.
These properties include the line-of-sight velocity distribution, the steepened size-linewidth relation, and the predominantly solenoidal nature of the turbulence. Within the simulation, these properties result from the Galactic shear in combination with the cloud's gravitational collapse. This is a strong indication that 
the Galactic gravitational potential
plays a crucial role in shaping the CMZ gas kinematics, and is a major contributor to suppressing the SFR by inducing predominantly solenoidal turbulent modes.
\end{abstract}




\section{Introduction}
\label{sec:intro}

The Central Molecular Zone (CMZ) is one of the most extreme star-forming environments in the Milky Way. The region contains a large reservoir of molecular gas ($\sim 10^7~\rm{M_{\odot}}$; \citealt{Dahmen1998}) within the innermost few hundred parsecs of the Galaxy, with temperatures ($\sim 100~\rm{K}$; \citealt{Ao2013,Ginsburg2016,Krieger2017}), column densities ($\sim 10^{23}~\rm{cm^{-2}}$; \citealt{Molinari2011}) and pressures ($P/k >10^7~\rm{K~cm^{-3}}$; \citealt{Rathborne2014,Walker2018,Myers2022}) much higher than in the Solar neighbourhood \citep{Kruijssen2013}. Despite that, the region as a whole has a star formation rate (SFR) which is an order of magnitude lower than expected based on the large amount of dense gas \citep[e.g. traced by NH$_3$;][]{Longmore2013}, and is likely due to a current minimum within an episodic star formation cycle \citep{Kruijssen2014,Armillotta2019,Callanan2021}. Sgr B2 accounts for at least 50\% of all star formation activity in the CMZ (possibly up to 89\%; \citealt{Barnes2017,Ginsburg2018}), leaving the rest of the clouds with quiescent to intermediate levels of star formation \citep{Lu2019,Walker2021,Williams2022}.

The interstellar medium (ISM) structure and star formation arise in response to the kinematic state of the gas \citep{Henshaw2020}. Therefore, the kinematics of the star-forming gas in the CMZ could help us understand the low SFR. The kinematics in the CMZ are also unususal, with high line-of-sight velocity dispersions and reports of a steep size-linewidth relation relative to the molecular clouds in the Galactic disk
\citep{Shetty2012,Kauffmann2017}. These phenomena are (at least partially) attributed to the effects of turbulence, which is known to play an important role in shaping the ISM \citep{Elmegreen2004,MacLow2004}. Turbulent motions consist of solenoidal and compressive modes that coexist at varied relative strength \citep[see e.g.][]{Federrath2010}. The compressive turbulent modes can lead to fragmentation and star formation by creating shocks and overdensities, while the solenoidal modes can prevent gravitational collapse. Within the CMZ we have an indication of predominantly solenoidal turbulence 
driving 
\citep{Federrath2016}, which is likely linked to the suppressed SFR. 
\citet{Orkisz2017} found an inverse relation between the fraction of solenoidal 
modes in the velocity field of 
the gas and SFR within Orion B. A later work by \citet{Rani2022} found the same type of relation for a large sample of Milky Way clouds at Galactocentric radii between $3{-}12$~kpc.

Even though turbulence is likely responsible for the kinematic and physical state of the CMZ clouds, it is currently not understood what drives it. Based on energetic analysis of common turbulence driving mechanisms, the CMZ turbulence is most likely driven by supernova feedback, followed by gas inflow from the galactic bar and magnetorotational instabilities \citep{Kruijssen2014,Henshaw2022}. However, this type of analysis is sensitive to coupling parameters that determine what fraction of the total energy goes into turbulent motions, and these parameters are not very well constrained. 
Recent work by \citet{Tassis2022} suggested that the CMZ turbulence can be explained by feedback from massive stars with high vertical (perpendicular to the Galactic plane) velocity dispersion that cross the clouds and deposit energy via stellar winds. The authors also demonstrated that this type of energy injection results in a steep size-linewidth relation.

An additional contribution  to the gas turbulence may come from
the strong orbital shear resulting from the Galactic gravitational potential \citep{Kruijssen2014,Krumholz2015,Federrath2016,Meidt2018,Keto2020}. This mechanism is expected to drive solenoidal turbulence within the gas, which is consistent with observational estimates \citep{Federrath2016}.

In this paper, we investigate how the Galactic gravitational potential affects the turbulence in the CMZ clouds. In particular, we focus on the G0.253+0.016 cloud, also known as `the Brick' \citep{Longmore2012}. This cloud is in the very early stages of star formation \citep[e.g.][]{Lis1994,Lu2019,Walker2021} and even though there is evidence that it may contain an H II region \citep{Henshaw2022a}, its kinematics are not dominated by in-situ stellar feedback. Furthermore, the Brick's structural and kinematic properties have been extensively studied through high resolution ALMA (Atacama Large Millimeter/submillimeter Array) observations \citep[e.g.][]{Rathborne2014,Rathborne2015,Federrath2016,Henshaw2019}. Here we use a recent cloud-scale hydrodynamics simulation \citep{Dale2019,Kruijssen2019, Petkova2023} which includes a model for the Galactic gravitational potential, and demonstrate that it matches very well the kinematic properties of the Brick. This analysis provides key predictions for the ongoing  ALMA CMZ Exploration Survey (ACES) on the Atacama Large Millimeter/submillimeter Array (Longmore et al.~in prep.), which will be able to characterise the driving mechanism(s) of turbulence in molecular clouds throughout the CMZ. 

\section{Simulation setup}
\label{sec:sim-setup}
We use the high-density (HDens) tidally-virialised simulation from \citet{Dale2019} (see their sect.~3 and tab.~1). \citet{Kruijssen2019} and \citet{Petkova2023} selected this particular model to represent the Brick as its initial conditions best match the cloud's size and mass. Furthermore, \citet{Kruijssen2019} showed that this simulation naturally reproduces other properties of the Brick, such as its column density and velocity dispersion (see their fig.~5). Additionally, \citet{Petkova2023} found similarities in the substrucure of the simulation and the real cloud in terms of their fractal dimension and spatial power spectra. Within this paper we expand the existing analysis of this simulation by performing a kinematic comparison to the Brick. In order to evaluate the importance of the initially assumed velocity field, we also repeat the analysis for the HDens self-virialised simulation from \citet{Dale2019} (see Appendix~\ref{sec:svir}). 

The simulation is performed with the smoothed particle hydrodynamics (SPH) code \textsc{gandalf} \citep{Hubber2018}. The simulation is three-dimensional, unmagnetised, and assumes an isothermal equation of state with temperature $65~{\rm K}$ \citep[consistent with the observed range for the Brick, e.g.][]{Ao2013,Ginsburg2016,Krieger2017} and a mean molecular weight $\mu=2.35$, corresponding to fully molecular gas. Self-gravity of the gas is included, whereas the field stars are included in the background potential (see below). The cloud is initialised as a sphere with total mass $\sim 4.5 \times 10^5~\rm{M}_{\odot}$ and $10^6$ SPH particles. The initial velocity field is turbulent with a power spectrum $P(k) \propto k^{-4}$, and virial parameter $\alpha_{\rm vir} = 3.2$. These initial conditions are selected from a set of randomly generated velocity fields to have negative spin angular momentum with respect to the orbital motion, consistent with the shear observed upstream from the Brick.

The simulated cloud is evolved on an eccentric orbit around the Galactic Centre starting $0.41~{\rm Myr}$ before the pericentre passage (see fig.~3 of \citealt{Kruijssen2019}) in the gravitational potential described in Appendix~A of \cite{Kruijssen2015}, which is based on the photometric model of \cite{Launhardt2002}. Since no turbulence driving is included, the initial turbulent velocity field of the cloud is quickly dissipated \citep[on a timescale $\approx 0.56~{\rm Myr}$;][]{Kruijssen2019}. Turbulence is generated during the simulation through gravitational collapse and shearing motions. Due to the lack of sufficient pressure support, the cloud fragments and forms sink particles (with threshold density of $\rho_{\rm sink}=10^{-17} \mathrm{g~cm}^{-3}$). By the time the present-day position of the Brick is reached (after 0.74~Myr of evolution), $\sim 55\%$ of the gas mass is transformed into sink particles.

For our analysis we focus mainly on the snapshot that corresponds to the present-day location of the Brick. We label this snapshot as being at $t=0~{\rm Myr}$. To facilitate analysis we bin SPH particles onto a 3D Cartesian grid with cell size $0.1~{\rm pc}$ using \textsc{splash} \citep{Price2007} and the exact mapping method of \citet{Petkova2018}. For reference, the sink accretion radius is $0.035~\mathrm{pc}$, and the median particle smoothing length is $0.096~\mathrm{pc}$. With the exception of Figures~\ref{fig:mom1} and~\ref{fig:size-lw}, which use the synthetic HNCO moment 1 map from \citet{Petkova2023}, all of the analysis is performed on these mapped simulation density outputs. The HNCO ($4_{04}-3_{03}$; 87.925~GHz) emission line is chosen, as within the Brick its emission is bright and extended, and it has been used in multiple observational studies \citep[e.g.][]{Federrath2016,Henshaw2019}.


\section{Comparison to the Brick}
\label{sec:brick-comparison}
\begin{figure}
	\includegraphics[width=\columnwidth]{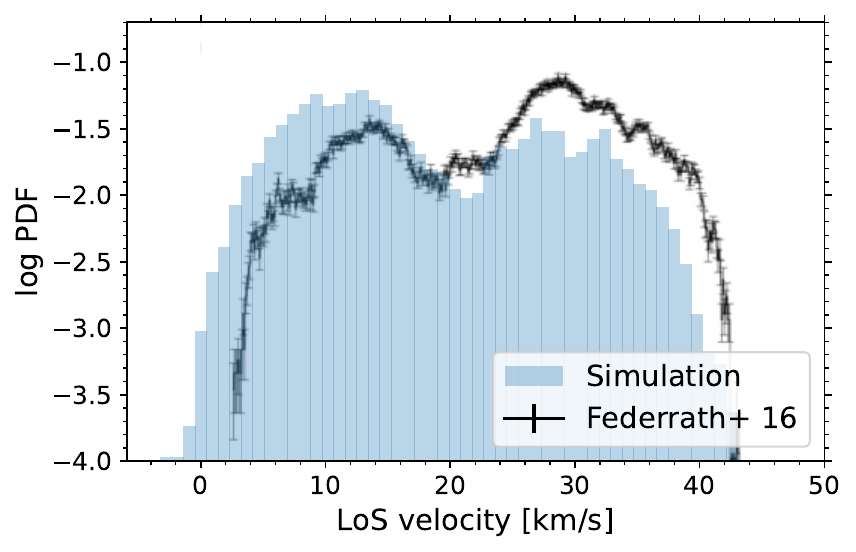}%
    \caption{The distribution of line-of-sight velocities in the first velocity moment map of HNCO ($4_{04}-3_{03}$) emission in the Brick. The blue histogram is obtained from 
    a similarly sized region from 
    synthetic observations \citep[][see their fig.~B1]{Petkova2023}. The black data points show the observed distribution in the Brick \citep{Federrath2016}.}
    \label{fig:mom1}
\end{figure}

In order to compare the kinematic state of the simulated and the observed cloud, we first consider their line-of-sight (LoS) velocities. \citet{Kruijssen2019} found that the simulation matches the LoS velocity dispersion of the real Brick, indicating a kinematic similarity between the clouds. In addition, the synthetic HNCO ($4_{04}-3_{03}$; 87.925~GHz) moment 1 map constructed by \citet{Petkova2023} shows a clear gradient and a matching LoS velocity range to the Brick (see their Appendix~B). Figure~\ref{fig:mom1} presents 
probability distribution function (PDF)
histograms of the synthetic moment 1 map and of the observed HNCO moment 1 map of the Brick \citep{Federrath2016}. The two distributions span the same velocity range and have a double-peaked profile, with a minimum at $\approx 20~{\rm km~s^{-1}}$. The results remain unchanged if we consider a synthetic moment 1 map that uses the density structure of the simulation instead of modelled HNCO emission. Note that both the spin angular momentum and the LoS velocity gradient of the simulation evolve with time (fig.~4 of \citealp{Kruijssen2019}), and the presented velocity distribution is not identical to the initial conditions. Furthermore, earlier simulation snapshots have very different LoS velocities. 

The double-peaked velocity profile in Figure~\ref{fig:mom1} is indicative of rotation along an axis perpendicular to the line-of-sight. However, the rotation is not necessarily global but it may be present in multiple structures within the Brick which are overlapping along the LoS \citep{Henshaw2019}. This is consistent with the velocity structure of the simulation, where the rotation is multi-axial, and broken down into spatially-coherent regions.

\begin{figure}
    \includegraphics[width=\columnwidth]{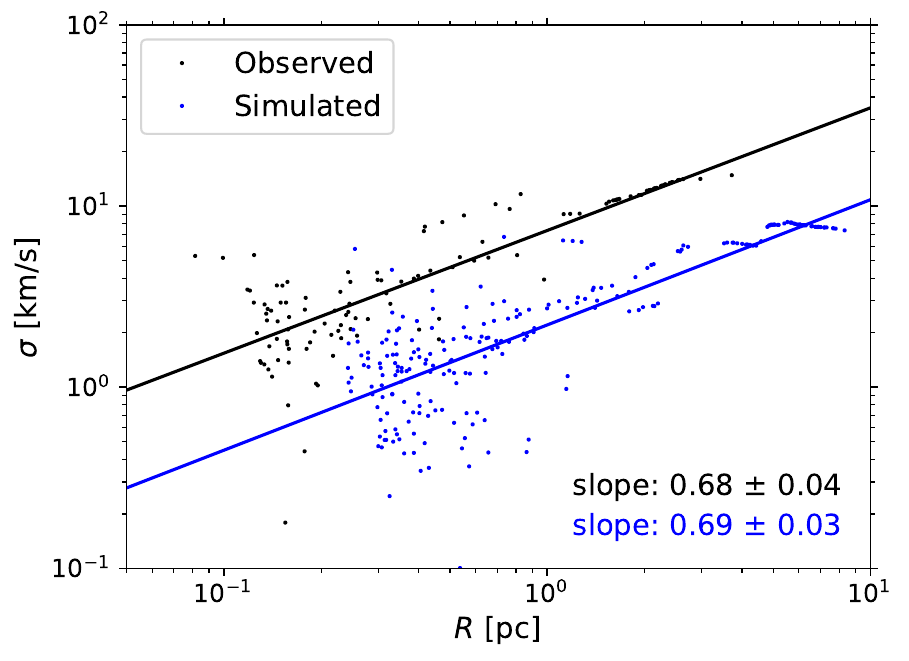}%
    \caption{Size-linewidth relation in the synthetic HNCO ($4_{04}-3_{03}$) emission map of the simulation snapshot \citep[data from][blue]{Petkova2023} and the Brick cloud \citep[data from][black]{Rathborne2015}. The individual data points correspond to structures identified within the corresponding PPV cubes using dendrograms. Power law fits for the two datasets are shown as solid lines, with both slopes being $\approx 0.7$.}
    \label{fig:size-lw}
\end{figure}

The LoS velocities can be used to construct the size-linewidth relation \citep{Larson1981}. We defer a full exploration of this observable in our simulations to a future study (Petkova et al. in prep.), but mention our finding that the simulated and observed cloud both exhibit the same size-linewidth slope ($\approx 0.7$; see Figure~\ref{fig:size-lw}). This is consistent with other CMZ studies \citep{Shetty2012,Kauffmann2017}, but is steeper than in the Solar neighbourhood \citep[$0.5$;][]{Heyer2015}. Our analysis 
considers the entire Brick cloud and
follows the procedure of \citet{Shetty2012}, which identifies structures in position-position-velocity (PPV) space with a dendrogram. For the simulation we construct a PPV cube using the HNCO ($4_{04}-3_{03}$) emission maps from \citet{Petkova2023}, and for the Brick we use the HNCO ($4_{04}-3_{03}$) PPV cube presented in \citet{Rathborne2015}. Figure~\ref{fig:size-lw} also shows a vertical offset between the two sets of data points, which can be explained as mismatch of pressure between the simulation and the Brick.

In contrast to the results shown in Figure~\ref{fig:size-lw}, \citet{Henshaw2020} performed a Gaussian decomposition of HNCO emission lines, and found a much shallower size-linewidth slope within identified sub-structures of the Brick. This suggests that the steeper relation may be due to rotational motions on the cloud scale.

\begin{figure}
    \includegraphics[width=\columnwidth]{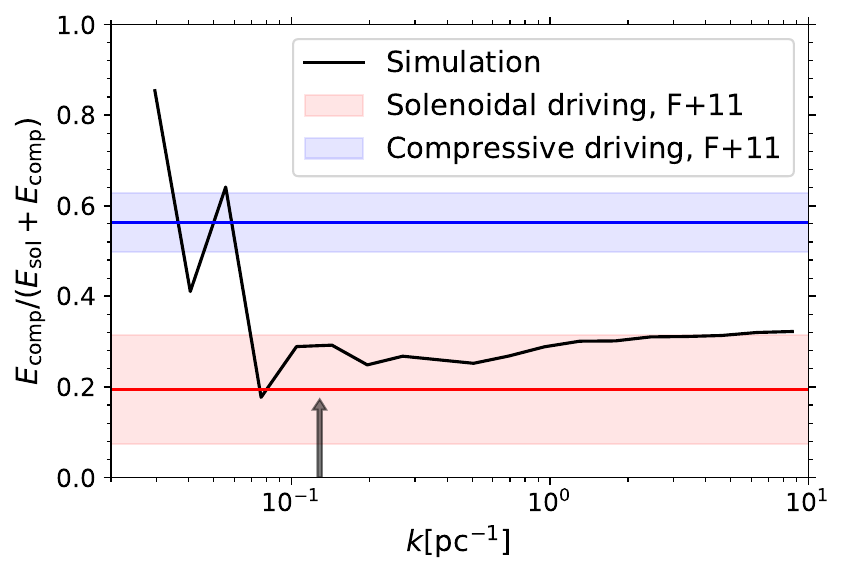}%
    \caption{Compressive ratio (kinetic energy in compressive modes of the turbulent velocity field divided by the total kinetic energy)
    as a function of spatial scale. The black line shows the ratio for our simulation, while the red and blue lines (and shaded areas) show the compressive ratio of simulations with purely solenoidal and compressive turbulence driving, respectively \citep{Federrath2011}. The 
    arrow indicates the (inverse of the) initial cloud size.}
    \label{fig:turb-driving}
\end{figure}

The similar (yet atypical) size-linewidth relation in the simulation and in the Brick is suggestive of a similar kinematic state, which is likely due to a combination of rotation and turbulence. \citet{Federrath2016} estimated the turbulence driving parameter of the Brick to be $b=0.22\pm 0.12$, which is consistent with having predominantly solenoidal driving. In order to compare this result with the simulation, we split the 3D velocity field into a compressive (curl-free) and a solenoidal (divergence-free) component using Helmholtz decomposition (see e.g. \citealt{Federrath2010}), and calculate the power spectrum of each component multiplied by the square root of the local density ($E_{\rm comp}$ and $E_{\rm sol}$, respectively). We then find the compressive ratio, $E_{\rm comp}/(E_{\rm comp}+E_{\rm sol})$, which represents the fraction of kinetic energy stored in the compressive modes of the velocity field. For supersonic clouds, the compressive ratio is always greater than 0, even if the driving force is purely solenoidal \citep{Federrath2010,Federrath2011}. Figure~\ref{fig:turb-driving} shows the compressive ratio of the simulation as a function of spatial scale ($k$), compared to the results of \citet{Federrath2011} for a Mach number of $\approx 11$. For most spatial scales our simulation has a cmpressive ratio of $0.2-0.3$, which is consistent with having predominantly solenoidal turbulence driving. This is also in agreement with the results of \citet{Federrath2016} for the Brick. Similar results are seen for earlier simulation snapshots.

All of the above measurements are consistent with the hypothesis that the Galactic shear is influencing the cloud kinematics. We explore this hypothesis further in the following section.

\section{The role of the Galactic potential}
\label{sec:gal-pot}

\begin{figure*}
	\includegraphics[width=\textwidth]{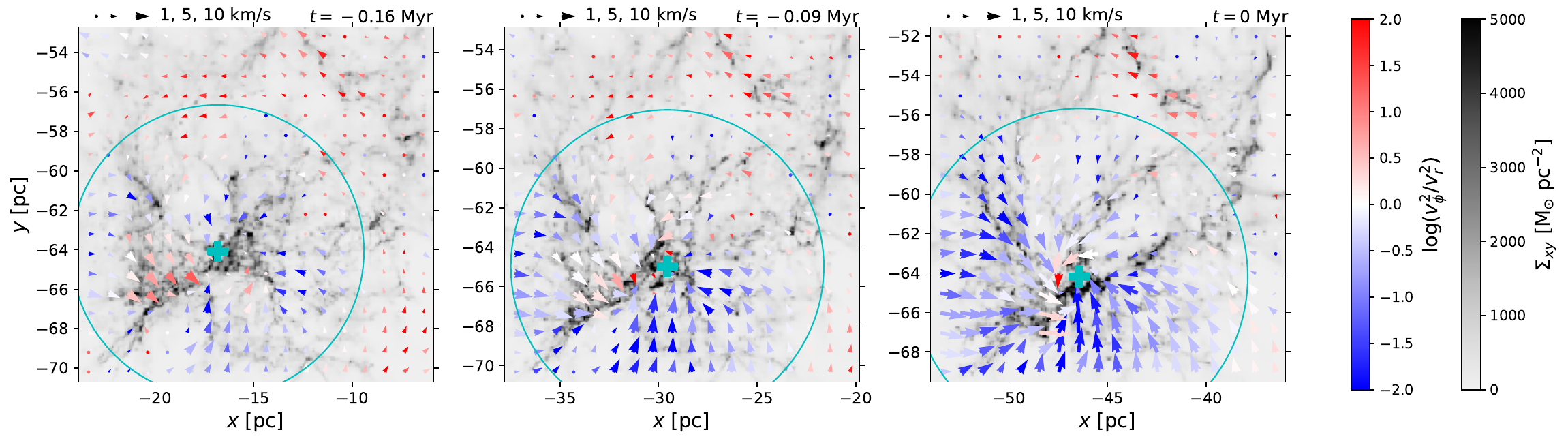}%
    \caption{Top-down view ($xy$-plane) of three snapshots of the simulated cloud (see time stamps). The column density is shown in grey scale, while the $xy$-velocities (mass-weighted averages along the $z$-axis) are shown as arrows. The length of each arrow indicates the magnitude of the corresponding velocity, with a $10~\rm{km}~\rm{s}^{-1}$ arrow drawn at the top of each panel for reference. Each arrow shows the velocity average within squares of $10 \times 10$ pixels. The cyan cross in each panel marks the location of the local minimum of the gravitational potential within the cloud, and the cyan circle shows the size of the tidal radius (see eq. \ref{eq:tidal-rad}) around the cyan cross. The arrows are coloured based on the ratio of azimuthal to radial kinetic energy with respect to the position of the cyan cross. In this coordinate system, Sgr A$^*$ is located at $(8.08, 0.00, -6.68)~{\rm pc}$, and an observer on Earth is looking along the $y$-axis (see \citealt{Dale2019}, fig.~2).}
    \label{fig:sim-rotation}
\end{figure*}

The Galactic gravitational potential can influence the evolution and dynamics of the CMZ clouds through two main effects: shear and tidal forces. The simulated cloud uses the \citet{Launhardt2002} potential, which has a scaling of $M\propto R^{2.2}$ between the enclosed mass $M$, and the Galactocentric radius $R$ for radii between 60~pc and 100~pc \citep{Kruijssen2015}. 
Using this dependence, \citet{Kruijssen2019} derived the velocity differential due to shear:
\begin{equation}
    \delta v_{\rm shear} = 0.67~\rm{km}~\rm{s}^{-1} \left ( \frac{\Omega_{\rm rot}}{1.7~\rm{Myr}^{-1}} \right ) \left ( \frac{\delta R}{1~\rm{pc}} \right ),
    \label{eq:shear}
\end{equation}
where $\Omega_{\rm rot}$ is the mean orbital angular velocity of a cloud (for our simulation $\Omega_{\rm rot}=1.7~\rm{Myr}^{-1}$; \citealp{Kruijssen2015}), and $\delta R$ is the difference in Galactocentric radius between two points in the cloud. While an updated potential \citep{Sormani2022} has been constructed since the simulation run, the shape of the new potential within the orbit of the simulation is consistent with that of \citet{Launhardt2002}, and hence the results of this paper remain unchanged.

The tidal radius of the cloud is \citep[][eq. 12.21]{Mo2010}:
\begin{equation}
    r_{\rm tidal} = \left ( \frac{m(r_{\rm tidal})/M(R)}{2 + \frac{\Omega_{\rm rot}^2 R^3}{G M(R)} - \frac{\mathrm{d}\ln M}{\mathrm{d}\ln R} \bigg\rvert_{R}} \right )^{1/3} R,
\end{equation}
where $m(r_{\rm tidal})$ is the cloud mass enclosed within the tidal radius. Note that $R$ is used for the Galactocentric radius and $r$ is used for the cloud-centric radius. By assuming that $\Omega_{\rm rot}^2 R^3/G M(R)=1$ (true for {circular} motion where $m \ll M$), and $\mathrm{d}\ln M /\mathrm{d}\ln R |_R=2.2$ \citep{Launhardt2002,Kruijssen2015}, we simplify the above expression to the following:
\begin{equation}
    r_{\rm tidal} = 5.36~\mathrm{pc} \left ( \frac{R}{70~\rm{pc}} \right ) \left ( \frac{m(r_{\rm tidal})}{10^5~\rm{M}_{\odot}} \right )^{1/3} \left ( \frac{M(R)}{2.8 \times 10^8~\rm{M}_{\odot}} \right )^{-1/3}.
    \label{eq:tidal-rad}
\end{equation}
In eq.~\ref{eq:tidal-rad} we express the dependence of the tidal radius on $m(r_{\rm tidal})$. This allows us to find $r_{\rm tidal}$ iteratively within the simulation. Note that due to the adopted gravitational potential, the tidal field is fully compressive \citep{Dale2019,Kruijssen2019}.

We now study the effects of shear and tidal forces on the kinematics of the simulation. Figure~\ref{fig:sim-rotation} shows a top-down view of the simulated cloud with superimposed $xy$-velocity vectors, where the bulk motion of the gas has been subtracted. We include three snapshots of the cloud -- one at the present location of the Brick (right), and two at earlier positions along the cloud's orbit. We find that as the cloud evolves it undergoes collapse towards a central dense region, which can be seen both in the more enhanced gas column density (grey scale in Figure~\ref{fig:sim-rotation}), and in the gas velocities. The velocity vectors are coloured based on the ratio of their tangential and radial components with respect to the local minimum of the gravitational potential along the orbit (cyan cross; hereafter `cloud centre').
Figure~\ref{fig:sim-rotation} shows that as the cloud evolves, there is more radial motion of the gas (blue arrows) concentrated within the tidal radius (cyan circle; see eq.~\ref{eq:tidal-rad}), and the regions outside the tidal radius move predominantly in a tangential direction (red arrows). This is consistent with the interpretation that the periphery of the cloud is
stretched due to shear, while its central region is collapsing (possibly with the help of tidal compression induced by the Galactic potential).

\begin{figure*}
    \includegraphics[width=\textwidth]{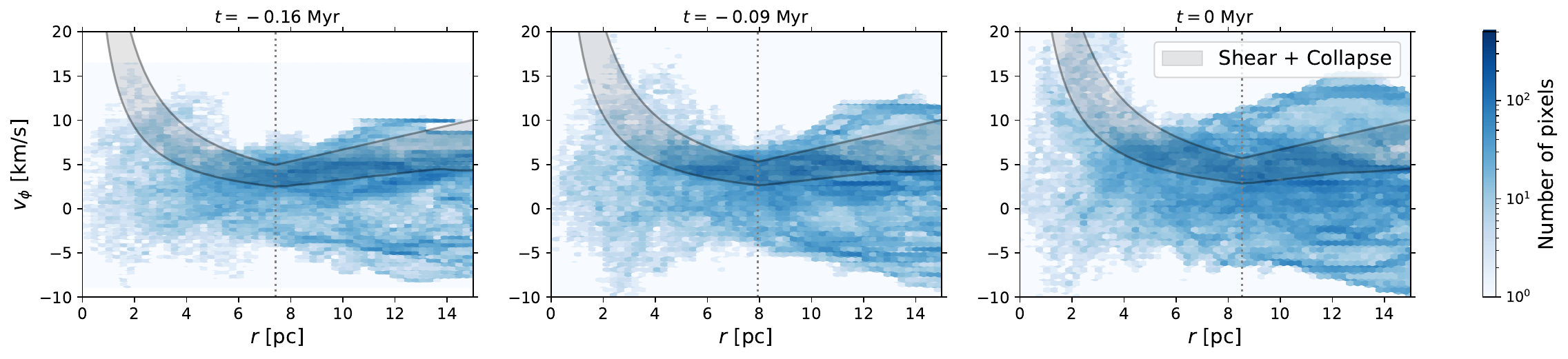}%
    \caption{Tangential velocity component as a function of radial distance from the cloud centre. The 2D histogram in blue presents the data from Figure~\ref{fig:sim-rotation}. The vertical dotted line marks the size of the tidal radius. The grey shaded area outside the tidal radius shows the expected tangential velocity based on shear (see Sec.~\ref{sec:gal-pot}). Inside the tidal radius, the boundaries of the grey shaded area follow $r^{-1}$ profiles, consistent with conservation of angular momentum during collapse.}
    \label{fig:shear-collapse}
\end{figure*}

In order to quantify the effect of the shear, we consider the tangential velocity components of the gas with respect to the cloud centre, $v_{\phi}$, and their dependence on the distance from this centre, $r$ (see Figure~\ref{fig:shear-collapse}). We also include the velocity ranges that we expect from a simple model of shear (outside the tidal radius) and collapse (inside the tidal radius). For the shear we consider two limiting cases. In the first case (lower estimate) we take each pixel from Figure~\ref{fig:sim-rotation} and we compute its shear velocity using eq.~\ref{eq:shear}. This approach does not give axisymmetric results with respect to the cloud centre. We then divide the pixels in radial distance bins and compute the mean $v_{\phi}$ in each bin. In the second case, we assume that a parcel of gas will maintain its tangential speed set by shear as the cloud rotates. This approach assumes that the effects of shear are effectively axisymmetric with respect to the cloud centre. To compute the upper velocity estimates, we use eq.~\ref{eq:shear} where we replace $\delta R$ with $r$. The grey shaded area is then continued within the tidal radius by assuming an $r^{-1}$ dependence for the upper and the lower velocity estimate. This is equivalent to a parcel of gas moving with the shear velocity at the tidal radius, and then being accreted while it conserves its angular momentum. 

Figure~\ref{fig:shear-collapse} shows that for all snapshots our lower theoretical prediction for the contribution of the shear (i.e.\ outside the tidal radius) overlaps with a prominent feature in the data. This feature is better defined in the early snapshots where the spread of velocities is smaller and there is less ongoing gravitational collapse. We also see an average increase of $v_{\phi}$ inside the tidal radius in all snapshots, consistent with spin-up due to collapse. This effect is most prominent at $t=0~\rm{Myr}$ where we have a better defined centre of cloud rotation. 

\section{Summary and Discussion}
\label{sec:discussion}
In this paper, we demonstrated that several kinematic properties of the CMZ cloud known as the Brick arise naturally in a hydrodynamics simulation which takes into account the Galactic gravitational potential. These properties include the line-of-sight velocity distribution, the steep slope of the size-linewidth relation and the solenoidally-driven turbulence. Within the simulation, we explain these through the effect of shear. In the outskirts of the simulated cloud, shear stretches the gas, boosts the velocity dispersion and seeds solenoidal turbulence. Due to the kinematic similarities between the simulation and the Brick, we conclude that the dynamical state of the Brick is likely strongly influenced by the Galactic gravitational potential. Our findings trigger several important follow-up questions.

\textbf{Can the turbulence be driven by another mechanism?} 
\textit{Within the simulation:} In addition to shear, turbulence can be driven by gravitational collapse within the cloud. \citet{Dale2019} compared clouds evolved with the Galactic potential to the same clouds evolved in isolation and found that the isolated clouds undergo more rapid collapse, but after the initial period of turbulent dissipation ($\approx 0.56$~Myr) their velocity dispersions remain lower than in the clouds evolved within the potential (see fig.~14 and~15 of \citealt{Dale2019}). Together with the solenoidal nature of the turbulence (see Figure~\ref{fig:turb-driving}), this indicates that the gravitational collapse on its own is not a sufficient turbulence driver. However, CMZ simulations which include the Galactic gravitational potential but no gas self-gravity also lack sufficient turbulence \citep{Hatchfield2021}. Therefore, the most likely interpretation is that shear seeds solenoidal turbulence which is amplified through gravitational collapse.
\textit{Within the Brick:} we cannot be sure that shear is the only factor contributing to the mode of the turbulence, but the agreement between simulations and observations suggest that it is likely to be an important factor. In addition to shearing motions within the cloud, there should also be shear with respect to the warmer diffuse gas surrounding the cloud, which can trigger Kelvin-Helmholtz instability. Other mechanisms can (and likely do) inject energy into the gas (e.g.\ stellar feedback; \citealt{Tassis2022,Henshaw2022a}), but this type of energy injection does not typically trigger solenoidal motions \citep{Menon2020}. 

\textbf{Is the Galactic potential suppressing star formation in the Brick?} Many authors have argued in favour of the Galactic shear as the mechanism responsible for suppressing star formation in the CMZ \citep{Kruijssen2014,Kruijssen2019,Krumholz2015,Federrath2016,Meidt2018,Meidt2020,Keto2020}. However, the SFR in our simulation ($\sim 0.3~\rm{M_{\odot}~yr^{-1}}$; \citealt{Dale2019}) is much higher than that of the Brick ($10^{-4}-10^{-3}~\rm{M_{\odot}~yr^{-1}}$; \citealt{Rathborne2014,Walker2021}). This discrepancy suggests that the low SFR in the Brick may be partially caused by physical factors missing from the simulation, such as magnetic and thermal support. Magnetic fields are known to delay star formation and prevent fragmentation. \citet{Petkova2023} found a difference in the width of the column density PDFs between the simulation and the Brick, which can be accounted for with the estimated turbulent plasma $\beta$ of the cloud \citep{Federrath2016}, indicating that magnetic fields are likely important for shaping the cloud structure. Additionally, the high gas temperature of the Brick is explained with shock heating \citep{Ginsburg2016}, as well as high levels of cosmic rays and interstellar radiation \citep{Clark2013}, that are not captured in our simulation.

Another reason for the different SFR in the simulation and the Brick may be the idealised simulation assumptions. The simulation was initialised as a gas sphere, which differs from the expected complex filamentary clouds that enter the CMZ \citep{Tress2020}. The assumed spherical initial state is unstable under the strong compressive tide in the vertical direction, and hence our simulation flattens rapidly. This vertical collapse may be artificially boosting the SFR, and the discrepancy with the Brick may be reduced by assuming more realistic initial conditions.
Furthermore, the simulated cloud exists in isolation and it is possible that the Brick has formed through gradual accretion of (higher kinetic energy) material, shifting the timeline of star formation to a later point along the Brick's orbit.

\textbf{Observational predictions.} The dust ridge of the CMZ consists of several predominantly quiescent clouds, of which the Brick is the most studied one. The analysis presented in this paper predicts that these clouds should also be strongly influenced by the shear induced by the Galactic gravitational potential. As a result, the clouds are expected to have predominantly solenoidal turbulent motions, steep size-linewidth relation, and kinematic signatures of counter-rotation. These predictions are based on the assumption that the clouds can be treated as isolated objects on a CMZ orbit. If we find discrepancies with the kinematic predictions, this could indicate an ongoing cloud assembly, or a form of cloud-cloud interaction.

As part of the ALMA
CMZ Exploration Survey (ACES), we have observed the full high column density ($> 10^{22}$~cm$^{-2}$) reservoir of the Galactic centre region at high spatial ($\sim 0.05$~pc) and spectral ($\sim 0.2$~km~s$^{-1}$) resolution (Longmore et al. in prep.). These data include the full dust ridge, and will be compared to the predictions of this work. In addition, ACES covers a lot of dense gas that has not been previously targeted by ALMA. The kinematic state and the three-dimensional geometry of this gas have not yet been studied, and the predictions included here can help constrain them.

\vspace{1mm}
Our analysis concludes that the dynamical state of the Brick is likely strongly influenced by the Galactic gravitational potential. These findings are extendable to the rest of the quiescent CMZ clouds and make predictions for their turbulent state.

\section*{Acknowledgements}
This work was carried out by the ACES Collaboration as part of the ALMA CMZ Exploration Survey.
MAP and JMDK acknowledge funding from the European Research Council (ERC) under the European Union's Horizon 2020 
(ERC Starting Grant MUSTANG; 714907).
MAP, JMDK,  SCOG and RSK acknowledge financial support from the Deutsche Forschungsgemeinschaft (DFG; German Research Foundation) via the collaborative research center (SFB 881, Project-ID 138713538) ``The Milky Way System'' (MAP, JMDK: subproj. B2; SCOG: subproj. A1, B1, B2, B8).
MAP acknowledges support from a Chalmers Cosmic Origins postdoctoral fellowship.
JMDK acknowledges funding from the DFG through an Emmy Noether Research Group (KR4801/1-1).
COOL Research DAO is a Decentralised Autonomous Organisation supporting research in astrophysics aimed at uncovering our cosmic origins.
SCOG and RSK acknowledge subsidies from the Heidelberg Cluster of Excellence STRUCTURES in the framework of Germany's Excellence Strategy (EXC-2181/1 - 390900948) and funding from the ERC via the ERC Synergy Grant ECOGAL (855130).
AG acknowledges support from the NSF under grants AST 2008101, 2206511, and CAREER 2142300, and from STSCI under grant JWST-GO-02221.001-A.
JDH gratefully acknowledges financial support from the Royal Society (University Research Fellowship).
VMR has received support from the project RYC2020-029387-I funded by MCIN/AEI /10.13039/501100011033.
C.F.~acknowledges funding by the Australian Research Council (Future Fellowship FT180100495 and Discovery Projects DP230102280), and the Australia-Germany Joint Research Cooperation Scheme (UA-DAAD).
L.C. acknowledges financial support through the Spanish grant PID2019-105552RB-C41 funded by MCIN/AEI/10.13039/501100011033.


\section*{Data Availability}

The data underlying this article will be shared on reasonable request to the corresponding author.



\bibliographystyle{mnras}
\bibliography{references} 




\appendix
\section{Self-virialised initial conditions}
\label{sec:svir}
We repeat the same kinematic comparison to the Brick presented in Sec.~\ref{sec:brick-comparison} using a snapshot of a different simulation from \citet{Dale2019}. The chosen simulation also has the HDens setup \citep[][see their sect. 3 and tab. 1]{Dale2019}, but the initial velocity field is self-virialised instead of tidally-virialised. The difference between the two is that the tidally-virialised simulation has additional initial velocity support against the compressive tidal fields of the Galactic gravitaional potential.

\begin{figure}
	\includegraphics[width=\columnwidth]{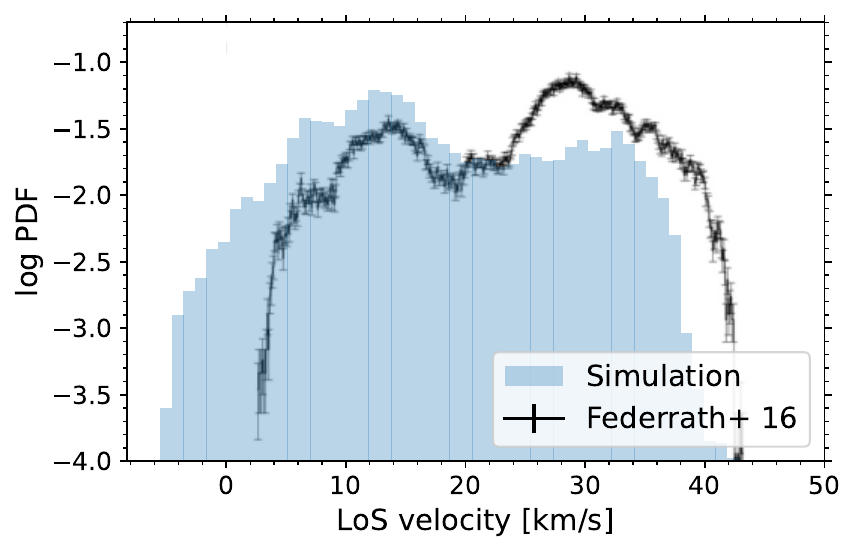}%
    \caption{Same as Figure~\ref{fig:mom1}, but using the self-virialised simulation snapshot.}
    \label{fig:mom1-sv}
\end{figure}

\begin{figure}
    \includegraphics[width=\columnwidth]{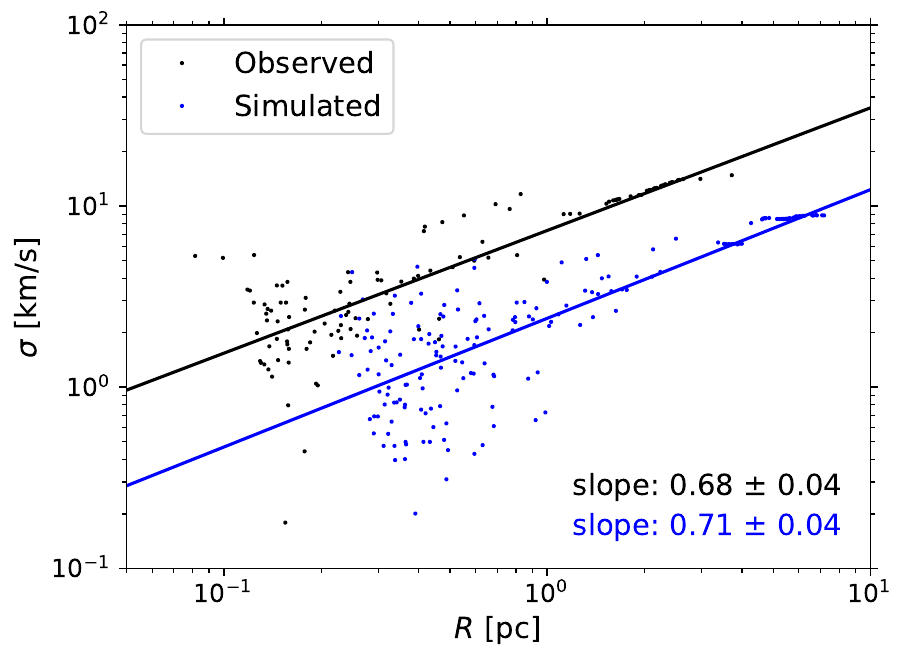}%
    \caption{Same as Figure~\ref{fig:size-lw}, but using the self-virialised simulation snapshot.}
    \label{fig:size-lw-sv}
\end{figure}

\begin{figure}
    \includegraphics[width=\columnwidth]{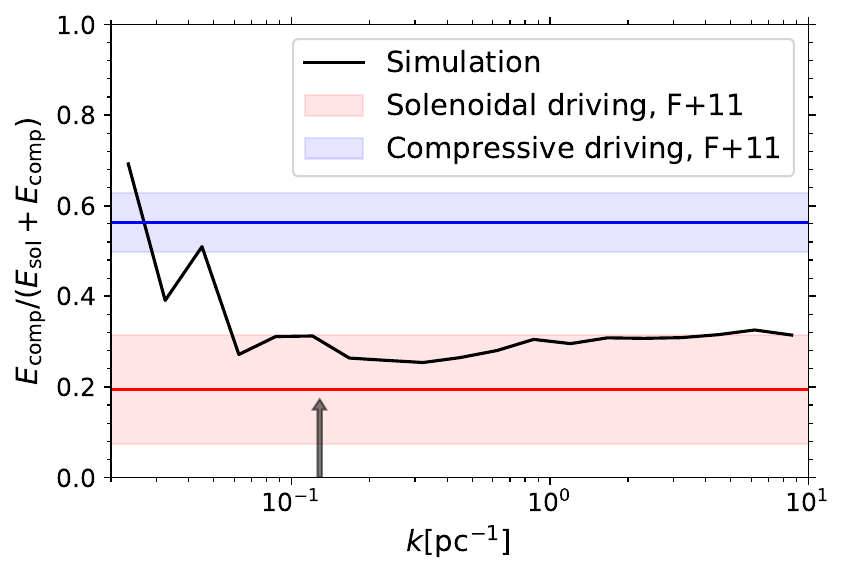}%
    \caption{Same as Figure~\ref{fig:turb-driving}, but using the self-virialised simulation snapshot.}
    \label{fig:turb-driving-sv}
\end{figure}

Figure~\ref{fig:mom1-sv}, \ref{fig:size-lw-sv} and \ref{fig:turb-driving-sv} collectively show that the main results presented in this paper hold for a simulation with a different initial velocity field. The line-of-sight velocity distribution is slightly less well matched to the Brick, but it shows a similar velocity range and a double-peaked profile about the same middle velocity value (Figure~\ref{fig:mom1-sv}). The simulated size-linewidth relation is similarly offset with respect to the observed one, with a slope which remains $\approx 0.7$ (Figure~\ref{fig:size-lw-sv}). And finally, the compressive ratio within the simulation is low and consistent with having predominantly solenoidal turbulence driving (Figure~\ref{fig:turb-driving-sv}).

\bsp	
\label{lastpage}
\end{document}